\documentclass[twocolumn]{aastex631}

\usepackage{listings}
\usepackage{CJK}

\def\SI#1#2{{#1}\,\mathrm{#2}}
\def\mailto#1{\href{mailto:#1}{#1}}

\def\china{China}

\shorttitle{FAST SETI Survey with A New Procedure}

\shortauthors{Wang et al.}

\begin{document}
\begin{CJK*}{UTF8}{gbsn}

\title{A Search for Extraterrestrial Technosignatures in Archival FAST Survey Data Using a New Procedure}

\correspondingauthor{Tong-Jie Zhang (\mailto{tjzhang@bnu.edu.cn}), 
Tingting Zhang (\mailto{101101964@seu.edu.cn}),
and Dan Werthimer (\mailto{danw@ssl.berkeley.edu})}

\author[0000-0002-8429-7088]{Yu-Chen Wang}
\affiliation{Institute for Frontiers in Astronomy and Astrophysics, Beijing Normal University, Beijing 102206, \china; \mailto{tjzhang@bnu.edu.cn}}
\affiliation{Kavli Institute for Astronomy and Astrophysics, Peking University, Beijing 100871, \china}
\affiliation{Department of Astronomy, School of Physics, Peking University, Beijing 100871, \china}

\author[0000-0002-4683-5500]{Zhen-Zhao Tao}
\affiliation{Institute for Frontiers in Astronomy and Astrophysics, Beijing Normal University, Beijing 102206, \china; \mailto{tjzhang@bnu.edu.cn}}
\affiliation{Department of Astronomy, Beijing Normal University, Beijing 100875, \china}
\affiliation{Institute for Astronomical Science, Dezhou University, Dezhou 253023, \china}

\author[0000-0001-9294-0363]{Zhi-Song Zhang}
\affiliation{National Astronomical Observatories, Chinese Academy of Sciences, Beijing 100012, \china}

\author{Cheqiu Lyu}
\affiliation{Kavli Institute for Astronomy and Astrophysics, Peking University, Beijing 100871, \china}
\affiliation{Department of Astronomy, School of Physics, Peking University, Beijing 100871, \china}
\affiliation{Department of Astronomy, Beijing Normal University, Beijing 100875, \china}

\author{Tingting Zhang}
\affiliation{College of Command and Control Engineering, Army Engineering University, Nanjing 210017, \china; \mailto{101101964@seu.edu.cn}}

\author[0000-0002-3363-9965]{Tong-Jie Zhang (张同杰)}
\affiliation{Institute for Frontiers in Astronomy and Astrophysics, Beijing Normal University, Beijing 102206, \china; \mailto{tjzhang@bnu.edu.cn}}
\affiliation{Department of Astronomy, Beijing Normal University, Beijing 100875, \china}
\affiliation{Institute for Astronomical Science, Dezhou University, Dezhou 253023, \china}

\author{Dan Werthimer}
\affiliation{Breakthrough Listen, University of California Berkeley, Berkeley, CA 94720, USA; \mailto{danw@ssl.berkeley.edu}}
\affiliation{Space Sciences Laboratory, University of California Berkeley, Berkeley, CA 94720, USA}

\begin{abstract}

The ``search for extraterrestrial intelligence'' (SETI) commensal surveys aim to scan the sky to find possible technosignatures from the extraterrestrial intelligence (ETI).
The mitigation of radio frequency interference (RFI) is an important step, especially for the most sensitive Five-hundred-meter Aperture Spherical radio Telescope (FAST), which can detect more weak RFI. 
In this paper, 
we propose several new techniques for RFI mitigation, and
use our procedure to
search for ETI signals from the archival data of FAST's first SETI commensal survey. 
We detect the persistent narrowband RFI by 
setting a threshold of the signals' sky separation,
and detect the drifting RFI (and potentially other types of RFI) using the Hough transform.
We also use the 
clustering algorithms 
to remove more RFI and select candidates. 
The results of our procedure are compared to the earlier work on the same FAST data. 
We find that our methods, though relatively simpler in computation, remove
more RFI (99.9912\% compared to 99.9063\% in the earlier work), but preserve 
the
simulated ETI signals except those (5.1\%) severely affected by the RFI.
We also report more interesting candidate signals,
about a dozen of which are new candidates that are not previously reported. 
In addition, we find that the proposed Hough transform method, with suitable parameters, also has the potential to remove the broadband RFI. 
We conclude that our methods can effectively remove the vast majority of the RFI while preserving and finding the candidate signals that we are interested in.
\end{abstract}

\section{Introduction}

The search for extraterrestrial intelligence \citep[SETI;][]{Cocconi59, Tarter01ARAA} aims to answer one of the most profound questions: are we alone in the universe? Unlike other approaches 
that search for
biosignatures \citep[including but not limited to the products of biological processes; e.g.][]{Webster15, Roth14}, SETI searches for technosignatures of intelligent civilizations. Since 1960s \citep{Drake61}, SETI has mainly been carried out in radio observation, striving to find possible signatures of radio emission produced by civilizations that can communicate via electromagnetic signals \citep[e.g.][]{Lebofsky2019PASP..131l4505L, Sheikh20, Zhang2020a, Smith21, Gajjar21, Tao22, Ng2022AJ....164..205N, Luan23, Ma2023NatAs...7..492M}. Though no rigid evidence of extraterrestrial intelligence (ETI) signals has been confirmed so far, efforts to answer this profound question will not stop.

Though we cannot rule out the possibility of broadband ETI signals, most works on SETI radio observations mainly focus on searching narrowband signals. This is based on the fact that the narrowest known natural radio emission is $\sim\SI{500}{Hz}$ \citep[e.g.][]{Cohen87}, while narrowband radio signals are commonly used in human communications, easy to be distinguished from astrophysical sources, and can be produced with a relatively low energy \citep{Li20RAA}. 

There are generally two observation modes for the SETI radio observations, namely the commensal surveys and the targeted observations. Targeted observations focus on pre-selected objects, usually nearby stars; while the SETI commensal sky surveys observe large sky areas to find candidate ETI signals for later confirmation. For fixed single dish telescopes, commensal surveys usually use the ``drifting scan" observation mode, which utilizes the rotation of the Earth to scan in right ascension (RA). An example of SETI commensal surveys is the SERENDIP program \citep{Werthimer01SPIE}, which spent decades searching for narrowband ETI signals at the 305-meter Arecibo Observatory in Puerto Rico. Thanks to the expanding datasets of known exoplanets in recent years, much work has been done on targeted SETI observations 
\citep[e.g.][]{Sheikh20, Smith21, Gajjar21, Tao22, Luan23}.
Nevertheless, the SETI commensal surveys are still complements to the targeted observations, since commensal surveys have a few advantages: 1.~they search ETI signals in larger sky areas, 2.~have orders-of-magnitude longer observation time, and 3.~they are target-agnostic (and therefore might avoid anthropocentric biases in target selection).

As the largest single-aperture radio telescope so far, the Five-hundred-meter Aperture Spherical radio Telescope \citep[FAST;][]{Nan2006, Li2016} provides us with great opportunities to search for extraterrestrial technosignatures \citep{Li20RAA, Chen2021}. With its 19-beam receiver, FAST is one of the most sensitive and efficient telescope for the multibeam SETI observation. The first SETI observation with FAST was a drift-scan survey, observed during its commissioning in July 2019, and the preliminary results were reported in \citet{Zhang2020a}. The FAST's first targeted SETI observation has also been conducted towards 33 selected exoplanet systems \citep{Tao22, Luan23}. FAST will conduct more SETI observations in the future, both targeted searches and commensal surveys.

One of the challenging tasks in the data analysis of the SETI radio observations is the identification and mitigation of the radio frequency interference (RFI), signals generated by human rather than the ETI. Considering the large sensitivity of FAST, we can expect more weak RFI in its observational data. Though in common astronomical observations one can usually directly remove the narrowband RFI, in most SETI campaigns the expected signal morphology for the ETI signals is also narrowband. In the previous work \citep{Zhang2020a} analyzing the first SETI commensal survey of FAST, the Nebula\footnote{\url{http://setiathome.berkeley.edu/nebula}} and kNN pipeline were used to mitigate the RFI. Several types of RFI, i.e.~the ``zone RFI'', ``drifting RFI'' and ``multibeam RFI'' were removed by the pipeline, and the $k$-nearest neighbor (kNN) algorithm was also used to further mitigate the RFI. Though most RFI was successfully removed in \citet{Zhang2020a}, it is still meaningful to further improve the algorithms for the RFI mitigation as well as the candidate selection, e.g.~removing more RFI to reduce the work of visual inspection,
and finding more candidate for later investigation.

In this paper, we present a novel procedure of RFI mitigation for the SETI commensal multibeam radio survey that searches for narrowband ETI candidates. We propose removing the ``persistent narrowband RFI'' by rejecting frequency bins that contain signals distributing in a large sky area (larger than a threshold); we also propose detecting and removing the ``drifting RFI'' (i.e.~the narrowband RFI that drifts in frequency) by detecting lines on the time--frequency waterfall plot with the Hough transform method. We apply these methods to the same FAST SETI commensal survey data as analyzed in \citet{Zhang2020a}, and then used the kNN and candidate selection methods to complete the full procedure of RFI mitigation and candidate selection. 

The rest of this paper is organized as follows. We describe the RFI removal methods proposed and used in this paper in detail in Section \ref{sec:method}, then briefly introduce the data and report the results of RFI removal and candidate selection in Section \ref{sec:data_and_results}. In Section \ref{sec:pixel_size}, we explore and discuss more possibilities of our method based on the Hough transform, when different parameters are set. We finally conclude and discuss the results in Section \ref{sec:concl}.

 \section{Methods for RFI Mitigation}
\label{sec:method}

In a SETI commensal survey, the original input dataset for RFI removal programs is the record of ``hits". A ``hit'' here refers to the information about a potentially interesting signal that has a high signal-to-noise ratio (SNR) in its frequency channel at each moment (see Section \ref{sec:data} and \citet{Zhang2020a} for details). Each hit consists of the information including the time, frequency channel, telescope pointing\footnote{The offsets of the pointings of different beams are taken into account.}, etc., which can be used for detecting RFI. Most hits belong to the RFI, while some hits may belong to interesting candidates.
In contrast to targeted observations, we typically expect the commensal survey to extend over a very long period of time, and the recorded data are usually only these hits instead of the complete spectra at each moment.
Therefore, we need to analyze the hits data, which requires software different from those usually used for the ``filterbank'' data.

In this paper, the RFI is removed in three steps, namely ``persistent narrowband RFI removal'', ``drifting (narrowband) RFI removal'' and removal of RFI using the clustering algorithm.

\subsection{Persistent narrowband RFI removal}
\label{sec:method.persistent}
Although narrowband signals are searched in SETI programs, many RFI signals produced on Earth are also narrowband signals. The sources of these RFI include near-ground radars, television and radio broadcasts, artificial satellites, cell phone signals, etc. Unlike narrowband ETI signals, narrowband RFI signals in a specific frequency channel are usually observed in different sky areas (as they are not actually of astronomical origin), and can persist for a relatively long period of time. For the SETI commensal survey observations considered in this paper, the persistence of signals is to some extent equivalent to observing signals in a large sky area, since the telescope usually scans along RA due to the rotation of the Earth.

To remove this kind of RFI, we divide the whole frequency range (1000--1500 MHz for the FAST observation) into small bins, and define a threshold of angular separation on the sky. If the hits in a frequency bin are found to spread a sky area larger than the threshold (i.e.~the angular separation of at least one pair of hits is larger than the threshold), we call the signals in this frequency bin as affected by the persistent RFI, and remove all hits in the bin. The spirit of this method is similar to the ``on-off strategy" applied in targeted SETI observations, which observe both the target (on-source) and reference locations (off-sources), and reject RFI signals that are observed on both on-source and off-source locations.

In practice, we process the hits one by one in the order of time. For each frequency bin and each value of declination (Dec)\footnote{For a SETI commensal survey using the ``drifting scan'' mode, the Dec for each beam is constant, so there are only several possible values of Dec.}, we only record hits with the maximum or minimum RA. Whenever we are processing another hit in this frequency bin, we calculate the distances between this hit and all of the recorded hits in this frequency bin. Since the sky area observed in a reasonable time period is not too large, the maximum distance to the recorded hits is equal to the maximum distance to all previous hits in this bin. 
If the maximum angular distance is larger than the threshold, we mark this frequency bin as persistent narrowband RFI, and ignore all subsequent hits in this frequency bin.

The persistent narrowband RFI removed with the aforementioned method is to some extent similar to the zone RFI and multibeam RFI described in e.g.~\citet{Zhang2020a}. In \citet{Zhang2020a}, the zone RFI was defined as the frequency bins with numbers of hits larger than a threshold (set according to the Poisson statistics), and the multibeam RFI was defined as hits received at similar time and frequency but in nonadjacent beams (see \citet{Zhang2020a} for details). We note that a frequency bin with a great number of hits usually means a large angular separation on the sky, and hits
received by nonadjacent beams have relatively large angular separations.

Thus, our method in this part mainly deals with generally similar (though not identical) types of RFI as the zone/multibeam RFI defined in \citet{Zhang2020a}. However, we only record and compare several RA and Dec coordinates for each frequency bin, and do not need to find hits with similar time and frequency for each given hit. 
As a result, our method is simpler, and usually needs less computation. We also note that our method processes hits sequentially in chronological order and removes a band immediately when a large sky separation is found. This means that the method can be potentially used for the real-time RFI detection, and do not need to wait for a long time until the accumulated numbers of hits are large enough for the Poisson statistics.

\begin{figure*}[tp]
\centering
\includegraphics[width=1\linewidth]{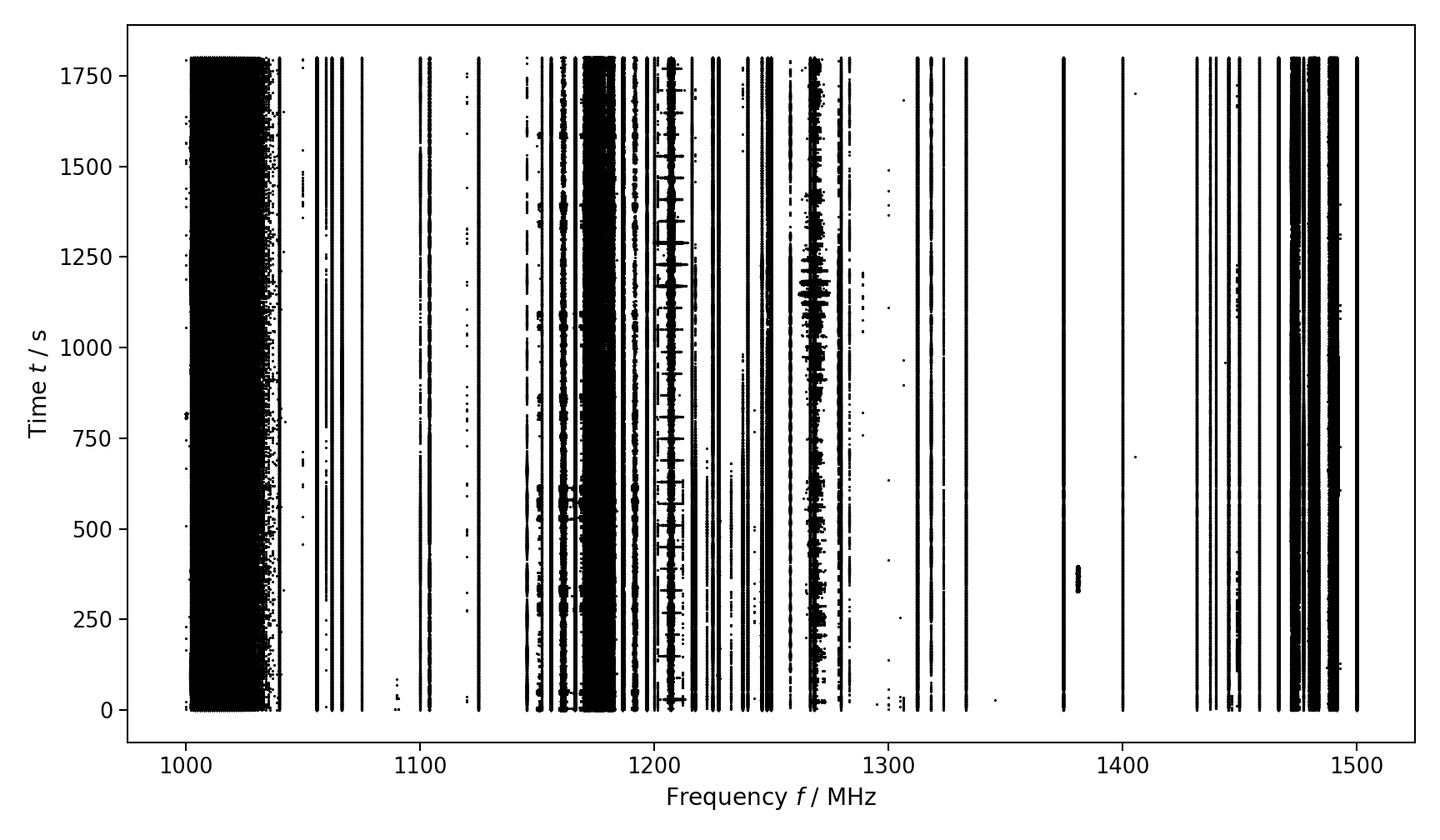}
\caption{The waterfall (frequency--time) plot showing the raw observational data (i.e.~``hits'' marked with black dots) in the first $\SI{1800}{s}$ of observation. The narrowband RFI is prominent, appearing as vertical lines; some broadband RFI (appearing as horizontal line segments) also exists. Throughout this paper, $t=0$ corresponds to $\mathrm{JD = 2458682.209155}$.}
\label{fig:raw1800s}
\end{figure*}

On a waterfall plot, i.e.~a plot of time $t$ versus frequency $f$ of the hits, a typical group of the persistent RFI is a vertical line, as can be seen in e.g.~Figure \ref{fig:raw1800s}.

\subsection{Hough transform for drifting RFI removal}
\label{sec:method.drift}
The drifting RFI is a special kind of narrowband RFI that drifts in frequency; some of the these signals are not fully understood. Possible origins include local oscillator malfunctions near the telescope, satellites, moving objects (e.g.~cell phones), etc. Since this kind of RFI drifts rapidly in frequency (relative to the frequency resolution), it can be missed when using the methods of rejecting frequency bins (like the persistent RFI removal method in this paper). To remove the drifting RFI, one needs approaches to detect this kind of drifting feature.

In SETI commensal surveys, drifting RFI removal methods can be designed based on the waterfall plot, i.e.~the time--frequency plane. 
In e.g.~\citet{Zhang2020a}, the drifting RFI is detected by defining symmetrical triangular bins for each hit and counting signals in the bins. If the number of signals in a triangular bin and its opposite three bins is above a threshold, the signals in these bins are defined as the drifting RFI (see e.g.~Figure 5 in \citet{Zhang2020a} for an illustration of this method). Similar methods are also described in e.g.~\citet{Cobb2000}.

In this paper, we 
propose
a method of removing the drifting RFI based on the Hough transform. As a commonly used method in the image analysis, the Hough transform can robustly detect straight lines or any parameterized curves in images. This method has been used to detect several kinds of signals in the time--frequency plane, e.g.~fast radio bursts \citep{Zuo2020} and sinusoidal ETI signals \citep{Monari2018}.
We note that the drifting RFI, although they may have no well-defined patterns, are also curves that can be clearly seen
and
detected with the Hough transform. Since short segments of a curve can be approximated as straight lines, we use the Hough transform to simply detect straight lines.

Hough transform detects patterns in images by ``voting" the parameters of a family of curves. For straight lines, the commonly used parameterization, suggested by \citet{Duda1972}, is
\begin{equation}
\rho = x\cos\theta + y\sin\theta,
\end{equation}
where $(x, y)$ are the coordinates of the points on a line, and $(\rho, \theta)$ are the normal parameters that specify the line. $\rho$ is the distance of the line from the origin, and the angle $\theta$ specifies the direction of the line's normal. Given a binary image, each (non-zero valued) figure point $(x_i, y_i)$ specifies a curve, 
\begin{equation}
\rho = x_i\cos\theta + y_i\sin\theta,
\label{eq:param_line}
\end{equation}
in the $(\rho, \theta)$ parameter space. After discretizing the parameter space into cells (usually $N_\rho\times N_\theta$ rectangular cells), Eq.~(\ref{eq:param_line}) gives a ``vote" of possible parameters in the $(\rho, \theta)$ space. Accumulating votes from all the figure points, one can find the local maxima of votes and get the parameters $(\rho, \theta)$ of detected lines.

To reduce the computation time, the probabilistic Hough transform was proposed \citep{Kiryati1991}. Rather than considering all the figure points, this method uses small subsets of figure points to detect patterns. In this paper, we use the \lstinline|OpenCV|\footnote{\url{https://opencv.org/}}
implementation of the probabilistic Hough transform, 
\lstinline|HoughLinesP|,
which is based on the progressive probabilistic Hough transform (PPHT) proposed by \citet{Matas2000}. The parameters of PPHT include accuracies of $\rho$ and $\theta$, the vote threshold (i.e.~the minimum vote of a detected line; in other words, the number of points on a line), the minimum length of a line and the maximum allowed gap in a single line. The algorithm can give the endpoints of all detected lines. For details, see \citet{Matas2000} and the \lstinline|OpenCV| documentation.

\begin{figure*}[t]
\centering
\includegraphics[width=\linewidth]{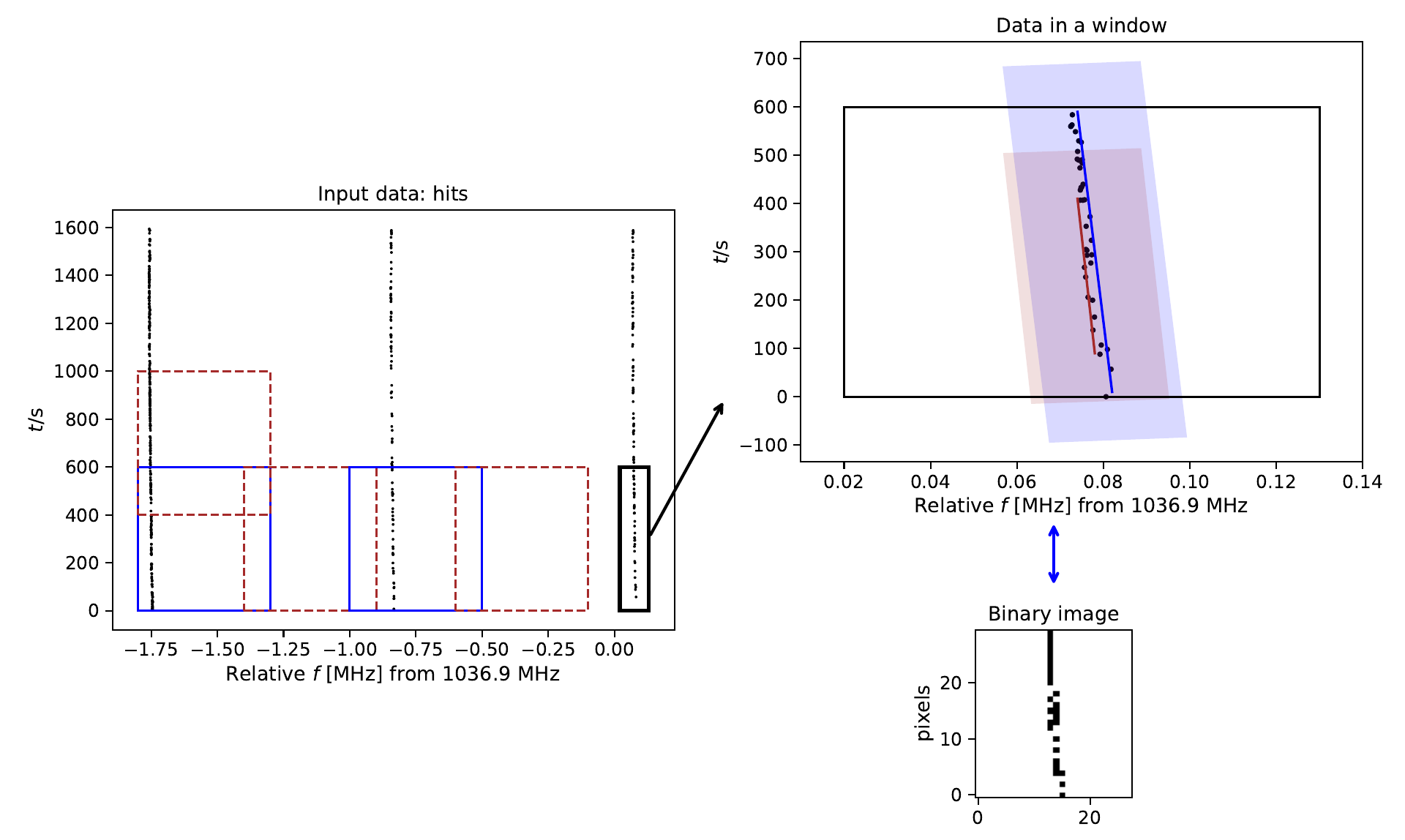}
\caption{Illustration of the drifting RFI removal method proposed in this paper. \textit{Left panel}: frequency $f$--time $t$ waterfall plot showing an example of several groups of drifting RFI, where the signals called ``hits'' are marked with black dots. Note that although the signals look like vertical lines, they are significantly drifting in frequency compared to the resolution, $\SI{3.725}{Hz}$. The $f$--$t$ plane is divided into windows with overlaps, as shown with blue and brown rectangles (different colors and line styles are used just for clarity). \textit{Right panels}: a small window that consists a group of drifting RFI for demonstration (also marked with the thick black rectangle in the left panel). The upper right panel shows the waterfall plot of the small window (the thick black box), which is converted to a binary image in the lower right panel. Lines are detected with the binary image using the Hough transform, and are shown in the upper right panel with the colored lines. Hits in a ``corridor'' around the detected lines, shown with the shaded regions, are marked as the drifting RFI. Note that all the parameters used to make this figure, except the width and overlap in frequency of the windows, are the same as those adopted in the implementation in Section \ref{sec:rfi_remove}.}
\label{fig:houghillu}
\end{figure*}

In this paper, we first set pixel sizes for frequency and time, and convert the scatter plot of ``hits''\footnote{As mentioned above, these hits refer to the recorded information of the high-SNR signals in the frequency channels at each moment, thus the input of the Hough transform is not the traditional waterfall plot of complete spectra (the matrix of signal powers at each frequency channel and time).} in the time--frequency plane into binary images, whose pixels in which there are hits are set to 1. To reduce the computation, the complete time--frequency plane is divided into an array of overlapping windows, and each window is converted into a binary image, as illustrated in Figure \ref{fig:houghillu}. For each image, line segments are detected with PPHT. Then, each line segment is extended by several pixels (so as not to miss the hits near the ends of the lines), and hits in a ``corridor" with a width of several pixels are marked as the drifting RFI. The process of removing drifting RFI using the Hough transform is illustrated in Figure \ref{fig:houghillu}.

\subsection{Clustering for RFI removal}
After removing the persistent RFI and the drifting RFI as described above, there are still RFI signals in the data. One example is the broadband RFI,
as shown in e.g.~Figure 10 in \citet{Zhang2020a}. This can be due to lightnings, power transmission cables near the surface of the Earth, electric fences, sparks, etc.
While the narrowband RFI tends to persist in a range of time but are restricted in a narrow range of frequency (though drifting may be present), the broadband RFI is characterized by transient surges in relatively wide frequency ranges but small time durations. This feature makes it difficult to detect broadband RFI with the aforementioned methods.

Following \citet{Zhang2020a}, we use the $k$-nearest neighbor (kNN) algorithm to remove more RFI from the data that pass the persistent and drifting RFI removal steps. Considering the nature of a SETI commensal survey, a set of ETI signals can neither persist in a long time duration (because of the drift of the telescope pointing), nor spread a wide frequency range (because of the commonly assumed narrowband nature). Thus, ETI signals would cluster on a scale smaller than that of RFI clusters. To remove large clusters that we consider as RFI, we calculate the mean distances to the $k$ nearest hits for each hit 
and those with a distance below a threshold are considered as RFI. The number $k$ should not be too small, so that hits in small clusters also have large mean distances, and only larger clusters are marked as RFI.

Though we mainly use the kNN algorithm for removing the broadband RFI (and other residual RFI), we also note that the Hough transform method described in Section \ref{sec:method.drift} has the potential to detect broadband RFI. We explore the details in Section \ref{sec:pixel_size}.

 \section{Data and Results}
\label{sec:data_and_results}

\subsection{The observed and simulated data}
\label{sec:data}

In this paper, we apply our method to the same observational data as in \citet{Zhang2020a}. The 5-hour data were collected during a drift-scan survey performed by FAST during commissioning in July, 2019.
The declination (Dec) of FAST's pointing was a constant,
but the accurate real-time pointing information of the FAST was not yet available during the time of that observation \citep[as discussed in Section 4.3 in][]{Zhang2020a}. To make it possible for us to remove the persistent narrowband RFI according to the sky separation (Section \ref{sec:method.persistent}), we approximately calculated RA and Dec for the hits of each beam according to the scan velocity. 
Although there can be errors relative to the real pointings, it is enough for the work of RFI removal.
We expect that accurate pointing information will be available in the future data of FAST's SETI commensal surveys.

The dataset of ``hits'' 
was generated from the results of a real-time SETI spectrometer, SERENDIP VI \citep{Cobb2000, Archer2016}, which was used to process the raw observational data. With the SERENDIP VI system, the power spectrum was calculated at each signal time, covering frequency bands from 1000--1500 MHz with a resolution of about 3.725 Hz. The power of each frequency channel, normalized with respect to the baseline, was compared to a signal-to-noise ratio (S/N) threshold. At each signal time, the frequency channels that exceed the threshold ($\mathrm{S/N}>30$) were recorded as ``hits". The information of each hit includes the signal power, time, frequency, telescope pointing at the moment, etc. Normally, only these hits, rather than the complete spectra, are recorded for commensal sky surveys, which is different from common targeted observations. The reason is that sky surveys are typically long-term observations, and recording the full spectra ($\sim 38$ billion spectral points per second) is expensive and difficult. For more information on the real-time data processing, see \citet{Zhang2020a}. 

The hits generated as described above are the initial data for a RFI removal program, and thus the input of our procedure. For reference, we plot the original data for the first $\SI{1800}{s}$, i.e.~the hits on the frequency--time ($f$--$t$) plane, in Figure \ref{fig:raw1800s}.

To check the methods proposed in this paper, we also add the same set of mock ETI signals, called ``birdies'', generated and described in \citet{Zhang2020a}. Each group of simulated signals was generated assuming a 
source, randomly located on the moving trajectory of 
FAST
and only has emission in one frequency channel. When a beam goes through the location, some hits are 
generated in the beam. As shown in Figure \ref{fig:candidate} (see also Figure \ref{fig:birdierfi} in this paper and 
Figure 13 in \citet{Zhang2020a}), there are a total of 20 simulated ``ETI signal sources'', generating 20 groups of birdies with 294 hits.

\subsection{RFI removal}
\label{sec:rfi_remove}

As described in Section \ref{sec:method}, our RFI removal procedure includes (1) persistent narrowband RFI removal, (2) drifting RFI removal and (3) RFI removal with the kNN algorithm. Since a hit can be part of the persistent narrowband RFI and the drifting RFI simultaneously, we only mark them as RFI rather than actually removing them in the first two steps. Then, we remove the two types of RFI together, after which the clustering analysis (step 3) is performed.

During the removal of the persistent narrowband RFI, we divide the frequency range ($\SI{1000}{MHz}$--$\SI{1500}{MHz}$) into bins with sizes of 7.45 Hz, twice the frequency resolution of the data.
A frequency bin with hits spreading over an angular distance of $0.14^\circ$ is marked as the persistent RFI. This threshold is $\sim 1.5$ times the distance between the centers of adjacent beams and about 2--3 times the beam size \citep[FWHM; as reported in e.g.][]{Jiang2019, Jiang2020}, so we can safely determine that the signals that exceed this threshold are in an extended area of the sky, and are unlikely to be ETI signals.  After this RFI removal step, 2,747,835 (4.094\%) of the frequency bins, consisting 99.7067\% of all the hits, are marked as RFI.

During the process of drifting RFI removal using the Hough transform, we divide the total frequency--time plane into windows with sizes of $\SI{20}{MHz}$ in frequency and $\SI{600}{s}$ in time. The overlaps of the windows are $\SI{1}{MHz}$ and $\SI{200}{s}$, and the pixel sizes when converting the hits in windows into images are $\SI{0.004}{MHz}$ and $\SI{20}{s}$. For the PPHT parameters (described in Section \ref{sec:method.drift}), we set the vote threshold to 5, minimum line length to 10 pixels, and maximum allowed gap to 8 pixels. Then, we extend the line segments by 5 pixels, and mark all hits with the distance less than 4 pixels from the line as RFI. As shown in Figure \ref{fig:houghillu}, these parameters are appropriate such that the typical drifting RFI can be resolved and detected with the Hough transform method. Note that all the parameters used to make Figure \ref{fig:houghillu}, except the width and overlap in frequency, are the same as the adopted parameters mentioned above. Since the frequencies of the RFI drift with scales much smaller than the width of the windows, changing the width and overlap in frequency has little effect on the RFI detection.
After this step, 99.5506\% of the hits are marked as drifting RFI.

\begin{figure*}[tp]
\centering
\includegraphics[width=1\linewidth]{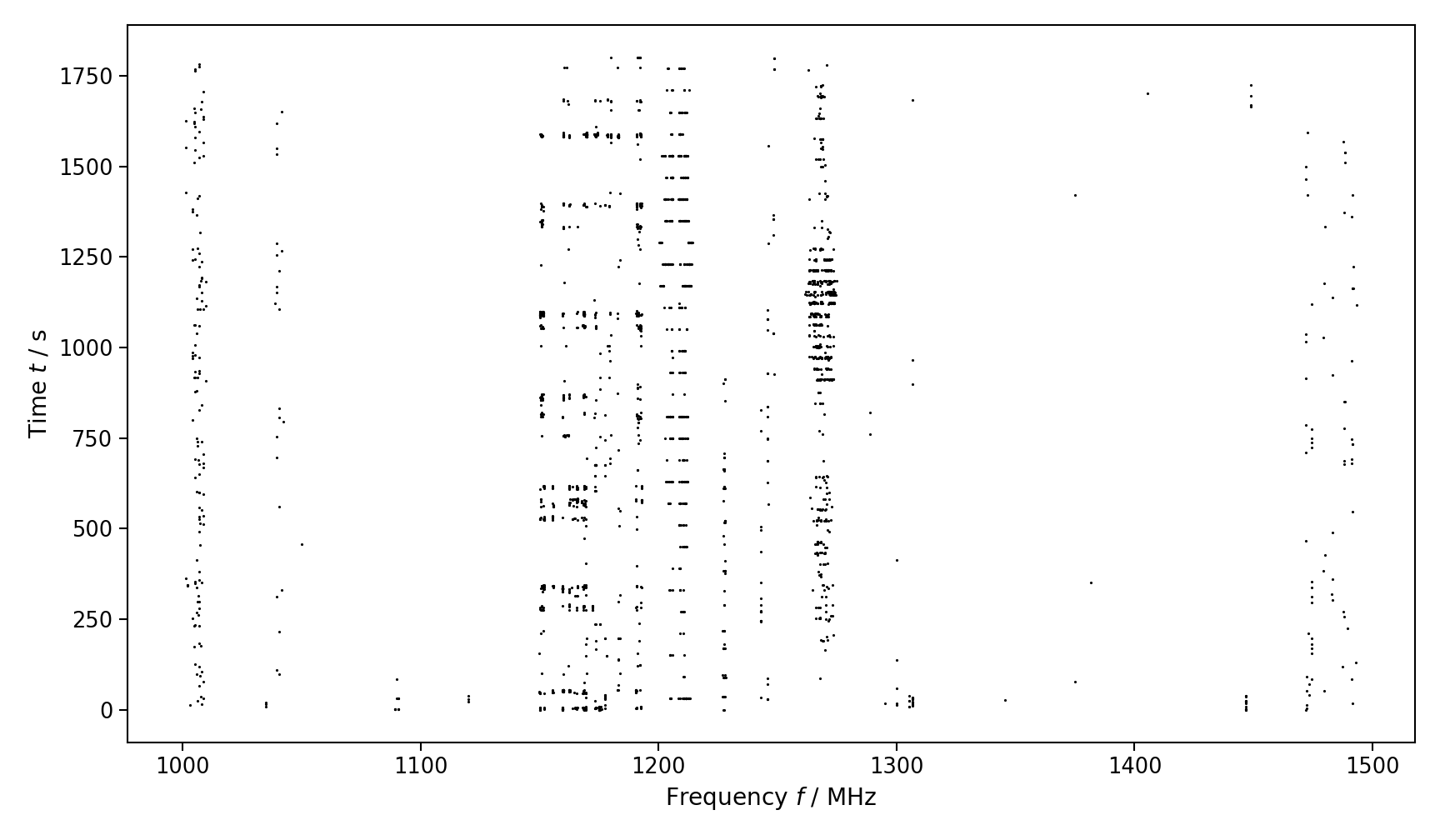}
\caption{Same as Figure \ref{fig:raw1800s} (with the same size of dots), but the hits marked as persistent or drifting narrowband RFI are removed. The narrowband RFI is greatly mitigated, though some broadband RFI still exists (and is mitigated later with the kNN method).}
\label{fig:perdrirfiremoved1800s}
\end{figure*}

After the first two steps, a total of 99.9912\% of the hits are removed as RFI signals, most of which are marked as both persistent RFI and drifting RFI; there are only 47,485 hits left for subsequent steps. 
The numbers of the two types of RFI signals and their ratios to all recorded signals are summarized in Table \ref{tab:rfi_ratio}. The data for the first $\SI{1800}{s}$ after the removal of persistent and drifting RFI is shown in Figure \ref{fig:perdrirfiremoved1800s}. As can be seen, the narrowband RFI is significantly mitigated by our method. By comparing to the results in \citet{Zhang2020a} (e.g.~98.1976\% zone RFI , 99.9063\% zone$+$drifting$+$multibeam RFI in Table 1 of \citet{Zhang2020a}), we find that we remove more RFI. We can also see in Figure \ref{fig:perdrirfiremoved1800s} that signatures of narrowband RFI is weaker, as compared to e.g.~Figure 14 in \citet{Zhang2020a}.

\begin{figure}[ht]
\centering
\includegraphics[width=0.95\linewidth]{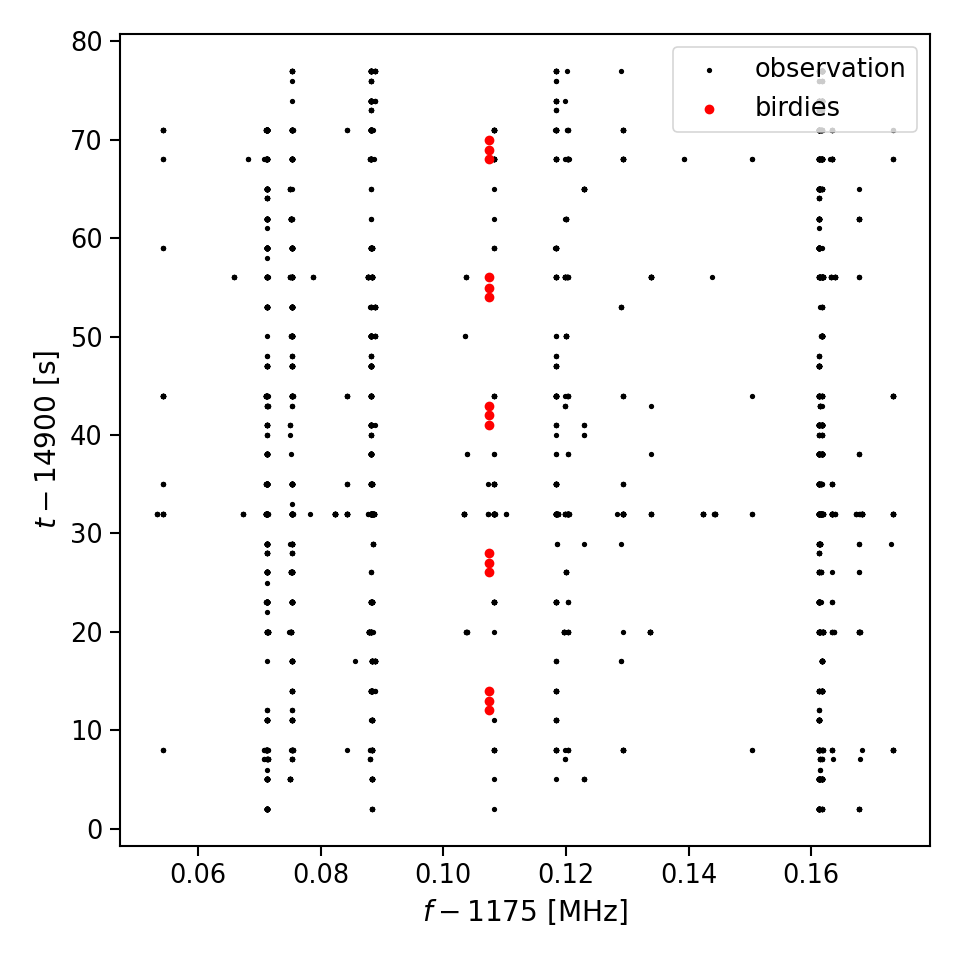}
\caption{A group of birdies (artificially simulated ETI signals) marked as drifting/persistent RFI is shown in red. The raw observational data are marked in black. These birdies are significantly affected by the RFI, so it is reasonable that the algorithms mark them as RFI.}
\label{fig:birdierfi}
\end{figure}

The birdies are added to the dataset before the RFI removal process, and only one group of 15 birdies (5.1020\% of the 294 birdies) are marked as (both) persistent and drifting RFI. 
As shown in Figure \ref{fig:birdierfi}, this group of birdies happens to be in an RFI-affected region (the frequency bin where this group is in is marked as RFI even without adding birdies). Thus, it is not surprising that the algorithm marks these birdies as RFI, since it is hard to distinguish birdies and RFI in this case. We can conclude that our method, with the parameters set above, removes most of the RFI while preserving all (mock) ETI signals except those severely affected by the RFI. 

\begin{table*}[tb]
\centering
\caption{Ratios of persistent and drifting RFI}
\label{tab:rfi_ratio}
\begin{tabular}{ccccc}
\hline\hline
& Persistent RFI & Drifting RFI & Both\footnote{Hits marked as both persistent and drifting RFI} & Total\footnote{Hits marked as any of the two types of RFI} \\
\hline
Number of hits & 538,956,414 & 538,112,874 & 536,574,702  & 540,494,586 \\
Ratio & 99.7067\% & 99.5506\% & 99.2660\% & 99.9912\% \\
\hline
\end{tabular} 
\end{table*}

\begin{figure*}[tp]
\centering
\includegraphics[width=0.9\linewidth]{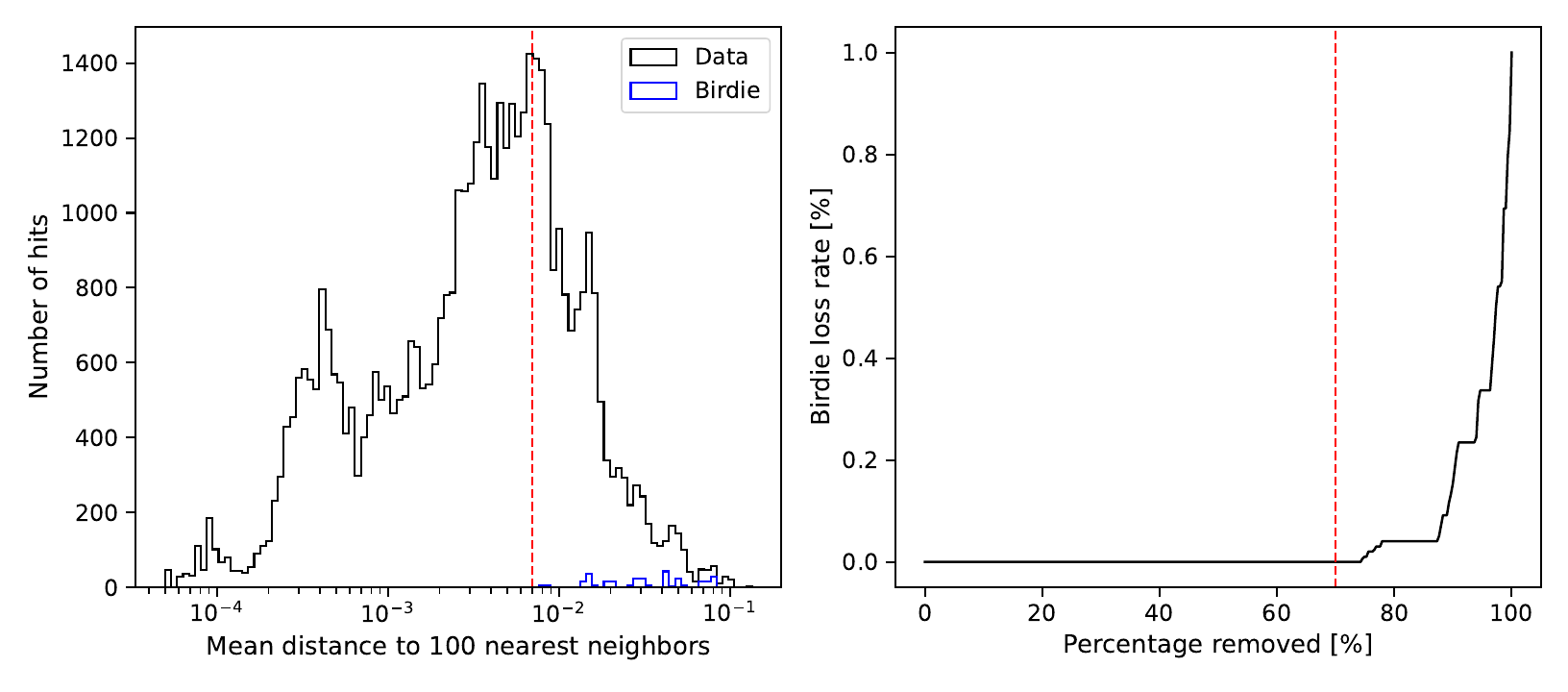}
\caption{The selection of the threshold for the RFI removal step with kNN, after removing the persistent and drifting RFI. \textit{Left panel}: Distribution of the mean distance between each hit and the 100 nearest neighbors on the normalized frequency--time plane. The distribution of all data (real observation and birdies) is shown in black, and the distribution for the birdies is shown in blue. The red vertical line shows the threshold for RFI removal (hits below it are removed). \textit{Right panel}: The birdie loss rate (the percentage of birdies that are removed as RFI) as a function of the percentage of all data removed (black curve). We set the threshold to remove 70\% of all data, as marked with the red vertical line.}
\label{fig:knndist}
\end{figure*}

\begin{figure*}[tp]
\centering
\includegraphics[width=0.9\linewidth]{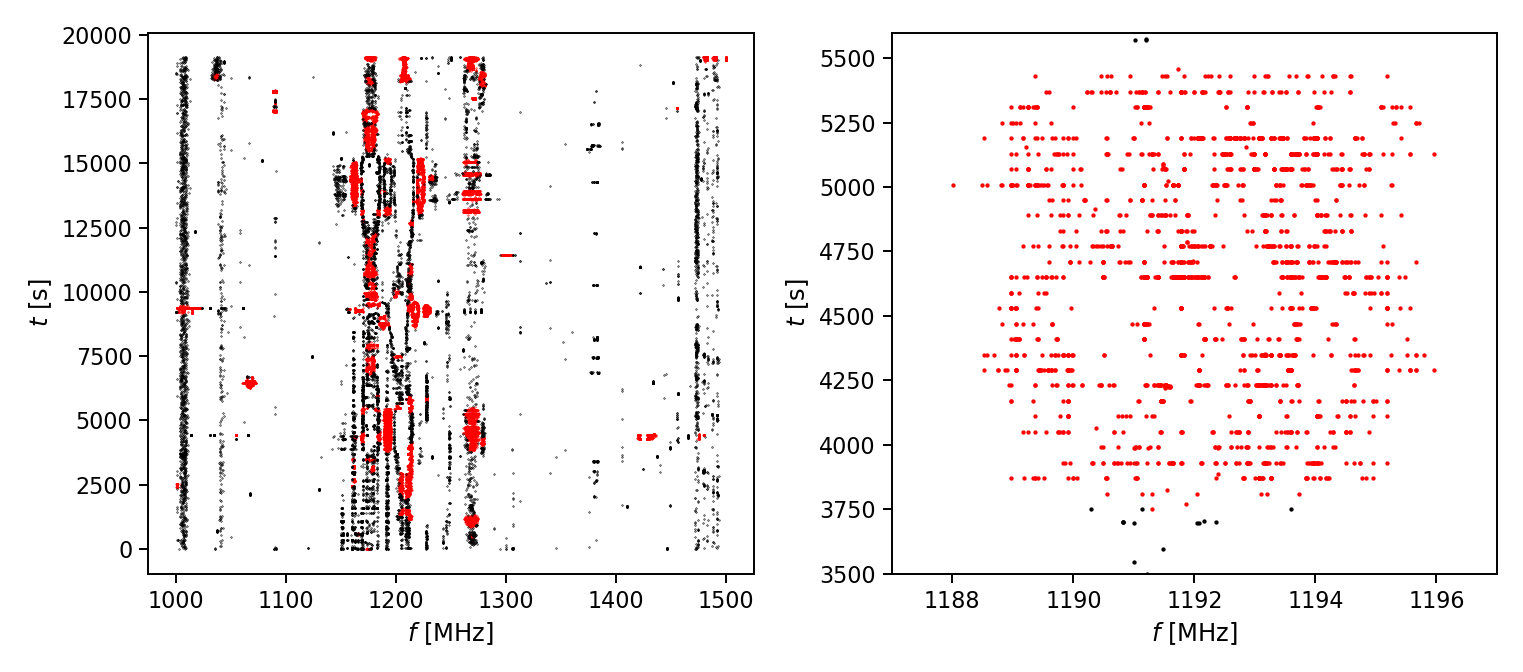}
\caption{The hits removed by the kNN are shown in red, and those not removed are shown in black. Left panel: All hits after removing the persistent and drifting RFI. Right panel: An example of a region where most hits of the broadband RFI are not removed before using kNN. The kNN effectively removes most of these hits.}
\label{fig:knnresultft}
\end{figure*}

The hits that pass the first two steps go through the last step of RFI removal based on the kNN algorithm. Following \citet{Zhang2020a}, we rescale the frequency and time of the data into the range of $[0, 1]$ (a common data preprocessing step for machine learning algorithms), and calculate the mean distance of each hit to the nearest $k=100$ (excluding itself) hits. The process is mainly implemented with the \lstinline|Python| package \lstinline|scikit-learn| \citep{scikit-learn}. As shown in Figure \ref{fig:knndist}, the mean distances for birdies are relatively larger than most observed hits. We set a threshold and remove 70\% of the hits whose mean distance are below the threshold, while no birdie is lost with this threshold. The hits removed with kNN are shown in Figure \ref{fig:knnresultft}. As can be seen on the right panel, some of the RFI removed in this step tends to be broadband RFI, which is hard to detect in the first two steps, since they mainly focus on the narrowband RFI.

In summary, our method of RFI removal effectively removes the vast majority of RFI, while preserving the simulated ETI signals (``birdies''). In the step of persistent and drifting narrowband RFI removal, a total of 99.9912\% hits are removed; in the kNN step, we can further remove 70\% of the remaining hits, many of which belong to the broadband RFI. We remove more hits than the result in \citet{Zhang2020a}, which removed RFI with 4 steps (zone, drifting, multi-beam RFI removal and the kNN method). Thus, in our test, our simple method removes more RFI without loss of more birdies.

\subsection{Candidate selection}
\label{sec:candidate}

After the three RFI removal steps in Section \ref{sec:rfi_remove}, we are left for only 14,040 hits. 
Though most hits left should still be RFI, the result is good enough for us to perform the subsequent candidate selection steps. Similar to \citet{Zhang2020a}, we assume that a group of meaningful candidate signals should be a small cluster of hits, and find the clusters using Density-based Spatial Clustering of Applications with Noise \citep[DBSCAN;][]{Ester96, Schubert17}.

\begin{figure}[htbp]
\centering
\includegraphics[width=0.95\linewidth]{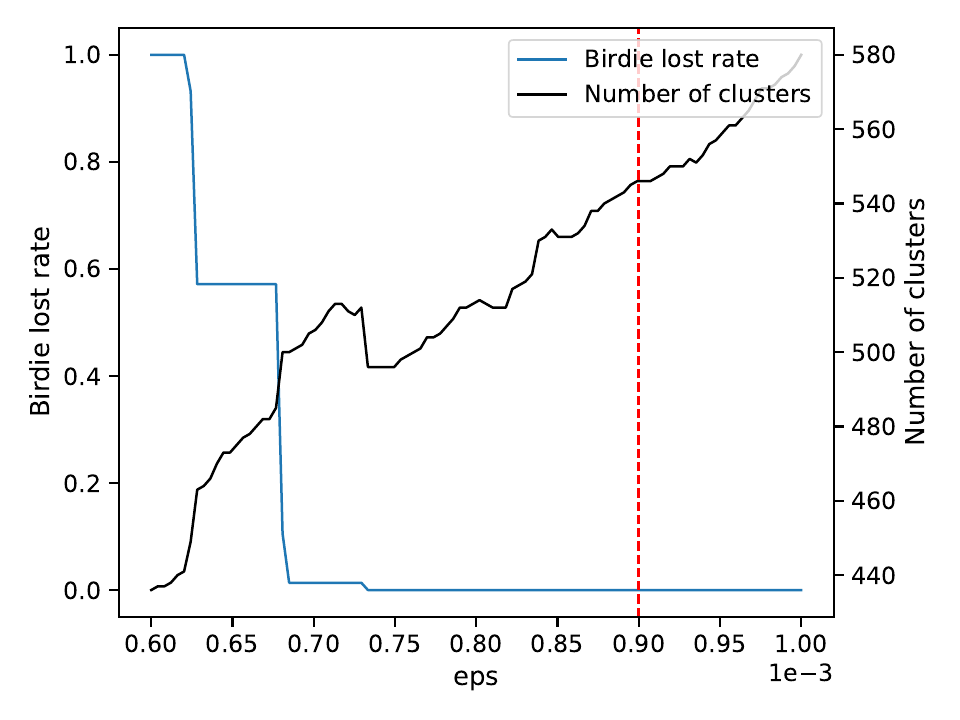}
\caption{The selection of the parameter Eps for the DBSCAN algorithm (red vertical line). The birdie loss rate (in blue) and the number of clusters (in black) detected with DBSCAN are also shown as functions of Eps.}
\label{fig:dbscancheck}
\end{figure}

The DBSCAN model detects clusters with two parameters (thresholds), \lstinline|minPts| and \lstinline|Eps|, as introduced in e.g.~\citet{Ester96, Schubert17}. It defines core points in the sample as those with more than \lstinline|minPts| neighbors within the radius of \lstinline|Eps|, and the neighbors that are not core points are called border points. A cluster is a group of core points and border points that are close to each other decided with the above thresholds, while the points that are neither core points nor border points are considered as noise. In our test, we rescale frequency and time to $[0, 1]$ (as for the kNN step), and set \lstinline|minPts|, \lstinline|Eps| to 5, $9\times10^{-4}$, respectively. These thresholds are set for the scaled values of frequency and time, and are chosen to preserve all of the birdies, as shown in Figure \ref{fig:dbscancheck}.

\begin{figure}[htbp]
\centering
\includegraphics[width=0.95\linewidth]{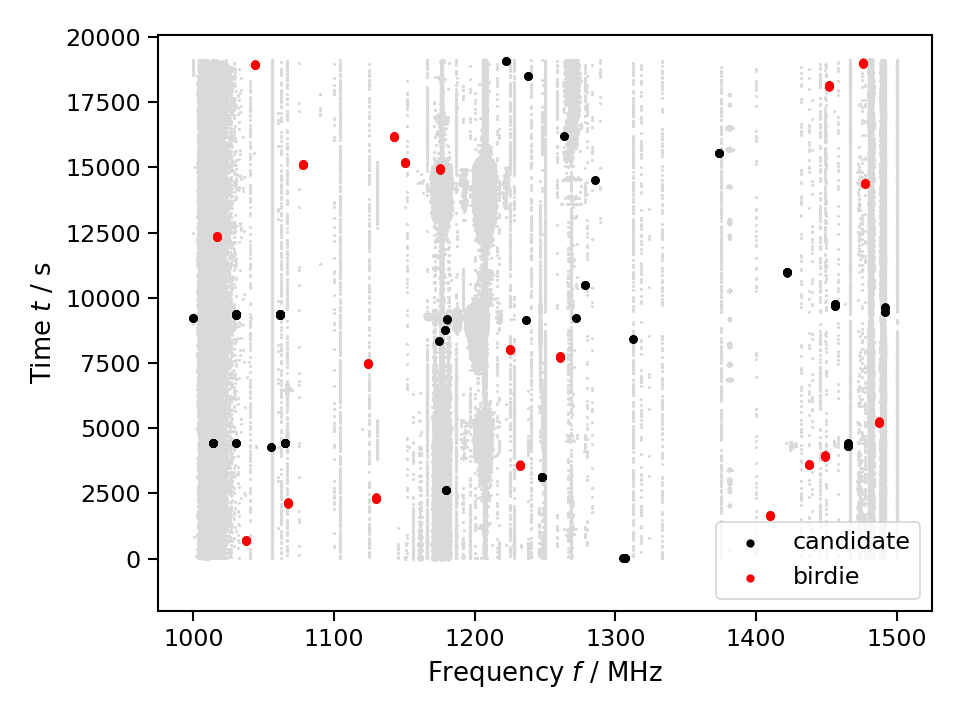}
\caption{The frequency--time plane showing candidate clusters which are detected by DBSCAN and pass the selection criteria of candidates. Candidates of real observational data are marked in black, and birdies are marked in red. The background shows a randomly drawn sample ($\sim 1/1000$ of all) of the raw observational data, since the raw data is too large to be plotted simultaneously.
Note that each dot consists of a cluster (or even several clusters) of hits, but the distances between them are too small to be resolved in this figure.}
\label{fig:candidate}
\end{figure}

With the above parameters, we use the DBSCAN implemented in \lstinline|scikit-learn| \citep{scikit-learn} and found 546 clusters (including 20 birdie clusters, which is consistent with the real number of birdie groups mentioned in Section \ref{sec:data}). We then select candidate clusters requiring that:
\begin{enumerate}
\item The maximum sky separation is less than 1.5 times the distance between the centers of adjacent beams, thus selecting clusters that do not distribute over a large area. (This automatically rejects clusters with signals simultaneously detected by non-adjacent beams, while allowing signals to be detected simultaneously by two adjacent beams or successively by different beams as a fixed point source on the sky drifts across the beams.);
\item The time duration is smaller than $\SI{100}{s}$ and the frequency bandwidth is smaller than $\SI{500}{Hz}$, thus selecting narrowband signals that do not persist for too long.
\end{enumerate}
All 20 birdie clusters satisfy the above criteria, and 31 candidate clusters of real data pass the selection, as shown in Figure \ref{fig:candidate}. We find fewer candidate clusters than \citet{Zhang2020a} (where 83 groups were selected), which is expected since we remove more RFI and fewer hits are left for the candidate selection.

We visually inspect the candidate clusters to check whether they have obvious features of RFI. As shown in \citet{Zhang2020a}, some selected clusters may actually be very close to other hits removed as RFI, and can be regarded as parts of the RFI that are missed in the previous steps. We extracted and checked the raw data within distances of $\SI{0.1}{MHz}$, $\SI{1000}{s}$ to the candidate groups, and preliminarily find $14$ promising groups  that do not seem to be parts of large clusters of RFI. 

\begin{figure*}[tp]
\centering
\includegraphics[width=0.99\linewidth]{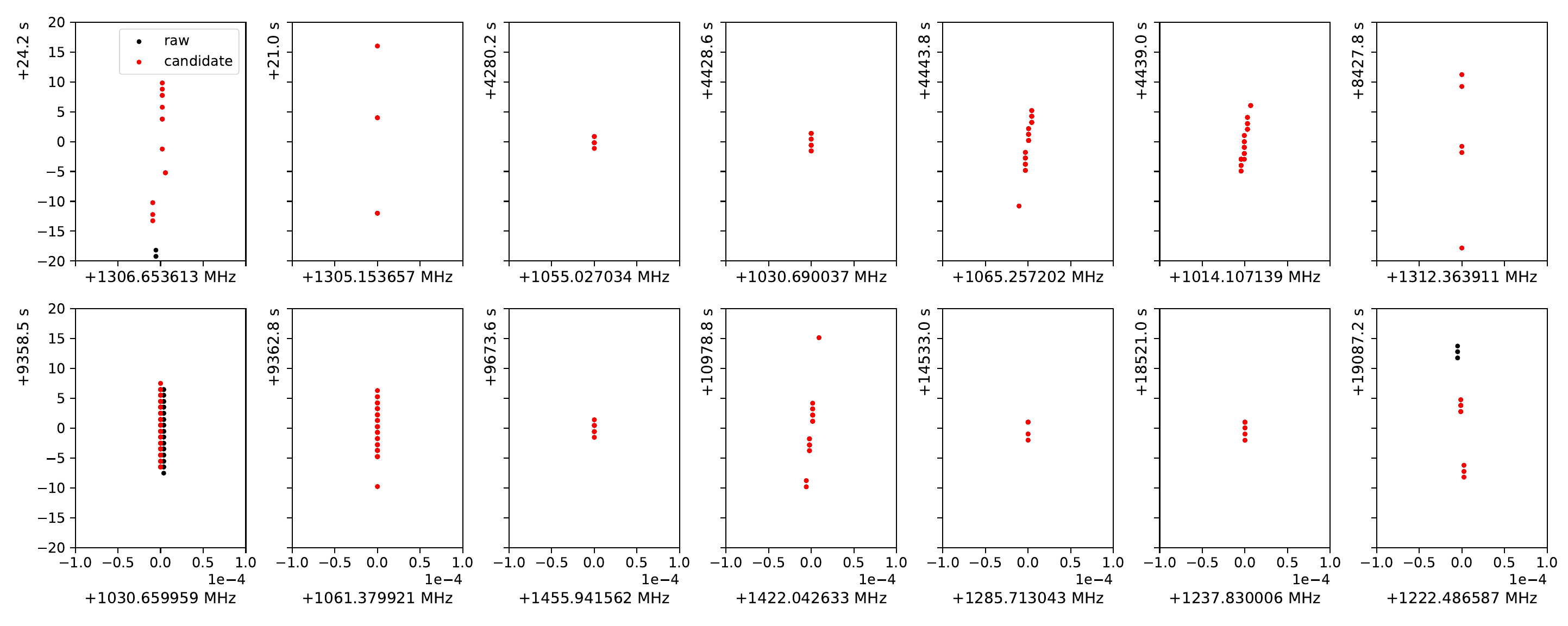}
\caption{Some interesting candidate groups found in this paper. The groups are detected with DBSCAN, selected with the criteria described in Section \ref{sec:candidate}, and no obvious evidence of large cluster of RFI is found near the candidate by visual inspection. The elements of the groups are shown in red, and other hits from the raw data that do not belong to the detected group are shown in black. All panels are shown in the same scale, and the positions of the centers of the panels are labeled on the axes.}
\label{fig:interesting}
\end{figure*}

The 14 interesting candidate groups are shown in Figure \ref{fig:interesting}. As can be seen, the candidates are indeed narrowband small clusters, which also look similar to the birdies. 
13 of the groups (all except that at $\sim(\SI{1055}{MHz},\SI{4280}{s})$) are newly found in this paper, as they are not reported by the previous work \citep{Zhang2020a}, which reported 2 groups of interesting candidates. Another group at $\sim(\SI{1055}{MHz},\SI{4430}{s})$ was also reported in \citet{Zhang2020a}, but is marked as RFI by the kNN step in this paper. As shown in the lower panel of Figure 18 of \citet{Zhang2020a}, this group of candidate belong to a relatively larger cluster. Thus, the hits in the cluster have smaller mean distances to the nearest neighbors, and are removed as RFI according to the threshold. 
This reminds us that the interesting candidates can still be parts of RFI that are missed during the removal procedure. Further verification and follow-up observations should be done before they can be considered as real ``signals of interest''.

By visual inspection, we also find that there are a few selected candidate groups that belong to the drifting narrowband RFI. Most of the cases are line segments of the drifting RFI where the points are so sparse that they may be missed by the Hough transform line detection. 
However, there is no candidate group that is part of a ``wide'' (simultaneously affecting several frequency channels at each moment) segment of the drifting RFI (as found in e.g.~Figure 16 in \citet{Zhang2020a}). This should result from the fact that we detect drifting RFI as hits within ``corridors'' (as described in Section \ref{sec:method.drift} and illustrated in Figure \ref{fig:houghillu}), rather than removing the hits in shapes like triangles (as in e.g.~\citet{Zhang2020a}). The triangle areas may miss some hits near the edge of the drifting RFI, which may later be detected by DBSCAN and pass the selection criteria, since the missed points can be very small clusters. On the other hand, our method tends to remove the drifting RFI more cleanly as long as the line segment of the drifting RFI can be detected.

In summary, we find about a dozen groups of interesting candidates that do not seem to be members of large clusters of RFI. \citet{Zhang2020a} found 2 groups of interesting candidates, and we report more new candidates that are very similar to the birdies, which represents the features of ETI signals that we are trying to detect. In addition, since we remove more RFI in the previous steps,
this leaves us with fewer candidate groups, reducing the work of the visual inspect.

\section{Discussion: Pixel Sizes for the Hough Transform}
\label{sec:pixel_size}

When removing the drifting RFI, we converted the scatter plots on the frequency--time plane into binary images, on which we performed the Hough transform to detect lines. The parameters of the process, especially the pixel sizes of the frequency and time, need to be reasonably chosen such that the RFI feature can be resolved in the images. As mentioned
in Section \ref{sec:rfi_remove}, we set the pixel sizes to $\SI{0.004}{MHz}$ and $\SI{20}{s}$ when detecting drifting RFI. The sizes are set considering the fact that, the drifting (narrowband) RFI tends to persist for a relatively long time, while drifting in a relatively small frequency scale, as shown in Figure \ref{fig:houghillu} (and e.g.~Figure 4 in \citet{Zhang2020a}). With these parameters, we effectively remove most RFI while preserving all birdies except those affected by RFI.

\begin{figure*}[tp]
\centering
\includegraphics[width=0.9\linewidth]{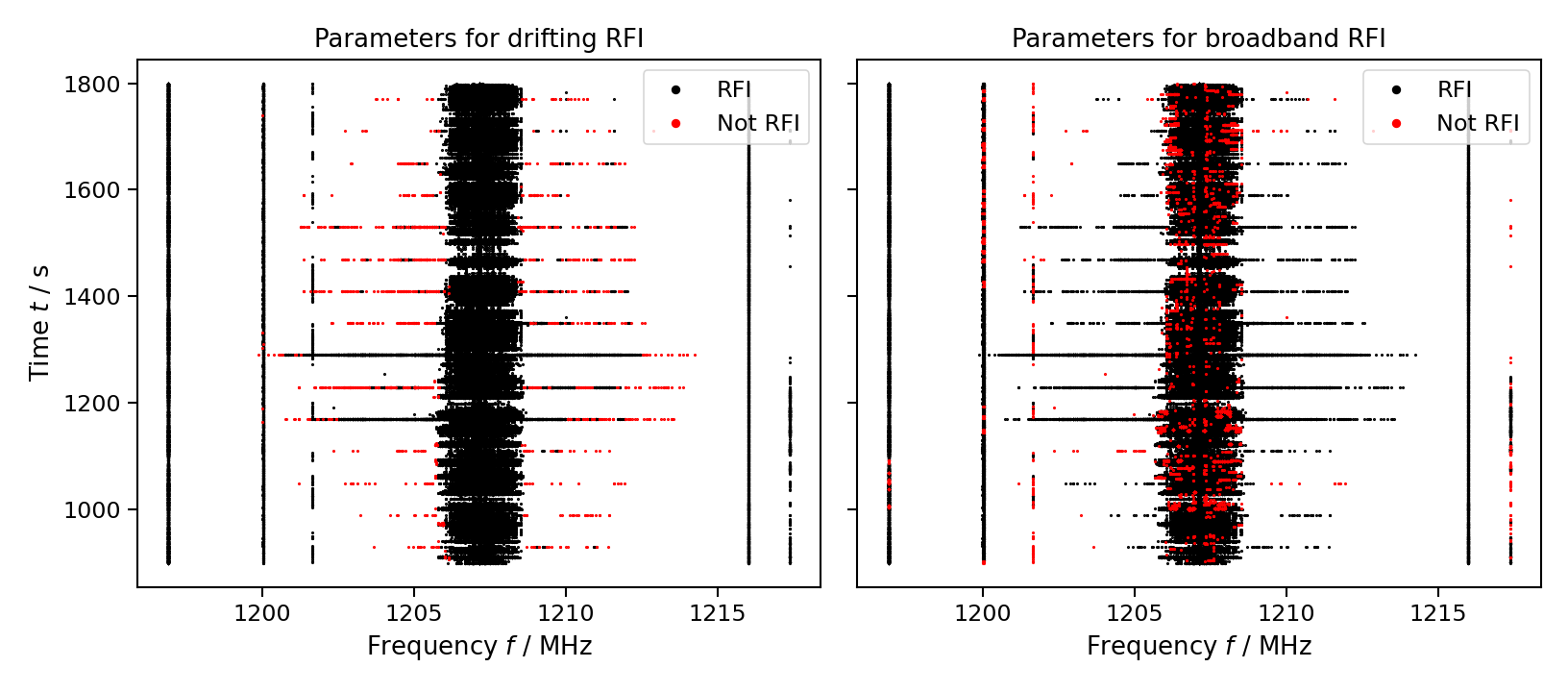}
\caption{An example area where both narrowband and broadband RFI can be seen, and the Hough transform is used to detect RFI, as described in Section \ref{sec:method.drift}. The hits detected as RFI are marked in black, and those not detected are marked in red. \textit{Left panel}: The parameters suitable of removing drifting RFI (pixel sizes are $\SI{0.004}{MHz}$ and $\SI{20}{s}$; as given in Section \ref{sec:rfi_remove}) are used. Most hits of the narrowband RFI are detected, while many of the broadband RFI are missed. \textit{Right panel}: The parameters suitable of removing the broadband RFI (pixel sizes: $\SI{0.06}{MHz}$ and $\SI{0.2}{s}$) are used. The broadband RFI is better detected, while some hits of the narrowband RFI are missed.}
\label{fig:broadbandexample}
\end{figure*}

However, as shown in the left panel of Figure \ref{fig:broadbandexample} (and also Figure \ref{fig:perdrirfiremoved1800s}), some broadband RFI is not detected by the drifting RFI removal method. A group of broadband RFI is typically a horizontal line on the frequency--time waterfall plot, which can be detected with Hough transform in principle. However, the frequency bands are so broad that, if the aforementioned pixel size $\SI{0.004}{MHz}$ is used, the hits within a group of broadband RFI are too far away from each other to be considered as a line.

We note that the Hough transform can also be used to detect the broadband RFI, provided that the pixels sizes (and other parameters, if necessary) are properly set to match the scale of the broadband RFI. We make a test using part of the raw data that contains both narrowband (persistent/drifting) and broadband RFI, as shown in Figure \ref{fig:broadbandexample}. By changing the pixel sizes to $\SI{0.06}{MHz}$ and $\SI{0.2}{s}$, our method with the Hough transform removes most of the broadband RFI, as shown on the right panel of Figure \ref{fig:broadbandexample}. However, with these parameters, some sparse narrowband RFI is missed, as expected. We find that the pixel sizes chosen for drifting RFI and broadband RFI complement each other.

In summary, though the Hough transform is mainly used to detect drifting (narrowband) RFI in this paper, we show that it is also capable of detecting the broadband RFI. However, broadband RFI appears to be horizontal lines on the $f$--$t$ plot, which are significantly different from the nearly vertical lines of the drifting RFI. Thus, different pixel sizes need to be set for the broadband RFI. Considering that the broadband RFI is much less dominant in the raw data, we used the parameters suitable for the drifting RFI in previous sections, and remove the undetected broadband RFI with kNN.

 \section{Conclusions}\label{sec:concl}

In this paper, we propose a new procedure of RFI mitigation and removal for the FAST multi-beam SETI commensal survey. Our methods remove the RFI in three steps, i.e.~the persistent narrowband RFI removal, drifting (narrowband) RFI removal and kNN RFI removal. By applying to the same FAST data and birdies analyzed in \citet{Zhang2020a}, we find that our methods can effectively remove the vast majority of the RFI while preserving the birdies, the simulated ETI signals. None of the birdies are detected as RFI, except for one group that is severely affected by the RFI. We detect and find a dozen of new interesting candidate groups, many of which look similar to birdies and do not obviously seem to be part of large RFI clusters. Thus, we conclude that our methods successfully mitigate the RFI for the SETI commensal survey, and help us find the signals similar to our simulated ETI signals. 

Compared to a previous work \citep{Zhang2020a} on the first SETI multi-beam observation with FAST (whose data is used to test our method), we use relatively simpler methods, remove more RFI in our test, and report some more interesting candidate groups. We use the simple threshold of sky separation (as called persistent RFI in this paper) to remove RFI similar to those removed as zone and multi-beam RFI in \citet{Zhang2020a}; for the drifting RFI, we use the Hough transform method to detect lines directly from windows consisting many hits and remove RFI signals in ``corridors'', rather than checking each hit for the number of other hits in triangular bins and removing hits in triangles. With these methods, we effectively removed the RFI, even more than the result reported in \citet{Zhang2020a}, and do not see more loss of birdies.

We also explored the effect of different parameters for the Hough transform method, especially the pixel sizes, on the performance of RFI removal. We find that, though mainly used for removing the drifting RFI in this paper, the Hough transform is actually capable of detecting the broadband RFI, provided that proper parameters are chosen. This suggests that our method based on the Hough transform is flexible, and has the potential of being applied to more tasks.

The RFI mitigation and candidate selection algorithms, either real-time or off-line, are important in the work of SETI surveys. In the future, we also plan to continue the study of improving these algorithms, utilizing the strength and characteristics of different methods. A better understanding of the properties of both RFI and ETI signals may help us improve the data analysis pipelines. For example, if we know the characteristics of some sources of RFI, it might be possible to search for candidate signals that are not consistent to the features of the known RFI, even in regions with a mixture of both ETI signals and RFI (as in e.g.~Figure \ref{fig:birdierfi}).

On the other hand, further studies on the properties of potential ETI signals are also useful. We note that the parameters of our methods, especially those for the kNN and DBSCAN methods, are set with reference to the birdies. Some parameters are chosen to make sure birdies are not removed by the algorithm. However, this process relies on a fully representative library of simulated ETI signals, including more possible patterns of birdies. For example, the pattern may be different for different brightness, drift rate \citep[e.g.][]{Li22} and the relative position to the beams. Some interesting candidates may be missed (removed with RFI) because there are no such kind of birdies that could guide our parameter selection. Thus, future work should make a more representative and diverse library of simulated ETI signals.

\begin{acknowledgments}
{We are grateful to the referee's insightful and useful comments, which helped us improve our manuscript. }
Tong-Jie Zhang (张同杰) dedicate this paper to the memory of his mother, Yu-Zhen Han (韩玉珍), who passed away three years ago (26 Aug.~2020).
We sincerely thank Pei Wang, Bo-Lun Huang, Jian-Kang Li for the useful discussions.  This work was supported by the National Science Foundation of China (Grants No.~61802428, 11929301). This work is finished on the workstation in Dezhou University.
\end{acknowledgments}

\software{Python, 
NumPy \citep{numpy},
pandas \citep{reback2020pandas, mckinney-proc-scipy-2010}, 
Astropy \citep{astropy},
Matplotlib \citep{Matplotlib},
scikit-learn \citep{scikit-learn},
OpenCV \citep{bradski2000opencv}.
}

\bibliographystyle{aasjournal}

\end{CJK*}
\end{document}